\documentclass[reprint,showpacs,preprintnumbers,nofootinbib,amsmath,amssymb,aps,twocolumn]{revtex4}
\usepackage{graphicx}% Include figure files
\usepackage{dcolumn}% Align table columns on decimal point
\usepackage{bm}% bold math
\usepackage{amsmath}
\usepackage{color}

\input{colordvi.tex}
\newcommand{\beq}{\begin{equation}}
\newcommand{\eeq}{\end{equation}}
\newcommand{\beqa}{\begin{eqnarray}}
\newcommand{\eeqa}{\end{eqnarray}}

\begin{document}

\begin{flushleft}
CTPU-PTC-18-02
\end{flushleft}

\title{Return of the grand unified theory baryogenesis: Source of helical hypermagnetic fields for the baryon asymmetry of the universe}

\author{Kohei Kamada}
\email[Email: ]{kkamada"at"ibs.re.kr}
\affiliation{Center for Theoretical Physics of the Universe,
Institute for Basic Science (IBS), Daejeon, 34126, Korea}
\affiliation{School of Earth and Space Exploration, Arizona State University, Tempe, Arizona 85287, USA}


\begin{abstract}

It has been considered that  baryogenesis models without a generation of $B$-$L$ 
asymmetry such as the GUT baryogenesis do not work 
since the asymmetry is washed out by the electroweak sphalerons. 
Here, we point out that helical hypermagnetic fields can be generated through the 
chiral magnetic effect with a chiral asymmetry generated in such baryogenesis models. 
The helical hypermagnetic fields then produce baryon asymmetry mainly at 
the electroweak symmetry breaking, which remains until today. 
Therefore, the baryogenesis models without $B$-$L$ asymmetry 
can still be the origin of the present baryon asymmetry. 
In particular, if it can produce chiral asymmetry mainly carried by  
right-handed electrons of order of $10^{-3}$ 
in terms of the chemical potential to temperature ratio, 
the resultant present-day baryon asymmetry can be consistent with our Universe, 
although simple realizations of the GUT baryogenesis are hard to satisfy the condition. 
We also argue the way to overcome the difficulty in the GUT baryogenesis. 
The intergalactic magnetic fields with $B_0\sim 10^{-16 \sim 17}$ G and $\lambda_0\sim 10^{-2\sim3}$ pc
are the smoking gun of the baryogenesis scenario as discussed before.

\end{abstract}
\maketitle
%\underline{\it Introduction}
\section{Introduction}
Baryon asymmetry of the Universe (BAU) is a long-standing problem in particle physics and cosmology. 
One of the popular models is the GUT baryogenesis~\cite{Yoshimura:1978ex}. 
The baryon asymmetry is provided from heavy boson decays
in grand unified theories (GUTs). 
The most troublesome issue is that only $B$ (baryon)+$L$ (lepton) but not $B$-$L$ asymmetry is generated 
in the heavy boson decay in the SU(5) GUT. 
The electroweak (EW) sphalerons~\cite{Kuzmin:1985mm} 
wash out the $B$+$L$ asymmetry and, hence, 
no asymmetry is left in that scenario. 

However, an interesting feature in the GUT baryogenesis, namely, 
the generation of chiral asymmetry, is still of interest. 
Since the first-generation electron Yukawa interaction, which is the weakest chirality flip interaction, 
is in equilibrium only at relatively low temperature, 
$T \lesssim 80 {\rm TeV}$~\cite{Campbell:1992jd}, 
the chirality is a good conserved quantity at higher energy scales. 
In particular, it has been noticed that with the help of the chiral magnetic effect~\cite{Vilenkin:1980fu}, 
maximally helical hypermagnetic fields (hyperMFs) are generated 
if sufficiently large chiral asymmetry exists in the thermal plasma at 
$T \gtrsim 80 {\rm TeV}$~\cite{Joyce:1997uy,Rogachevskii:2017uyc}. 
Such primordial helical hyperMFs can have a strong impact on cosmology~\cite{Vachaspati:2016xji}. 
They can be the seed for the galaxy and galaxy-cluster MFs and 
remain until today as the intergalactic MFs. 
Moreover, the baryon asymmetry is (re)generated through the 
Standard Model (SM) chiral anomaly~\cite{Giovannini:1997gp,Dvornikov:2012rk}, 
which is not completely washed out
by the EW sphalerons~\cite{Dvornikov:2012rk,Kamada:2016eeb}. 
Therefore, we can imagine the following scenario: 
(i) the GUT baryogenesis first generates the $B$+$L$ and chiral asymmetry, 
(ii) maximally helical hyperMFs are generated from the chiral asymmetry
while $B$+$L$ asymmetry is eventually damped, and 
(iii) the hyperMFs (re)produce baryon asymmetry, especially at 
the EW symmetry breaking, which lasts against the EW sphalerons. 
Then the asymmetry is responsible for the present Universe. 
In other words, the GUT baryogenesis can be the indirect origin of the present BAU. 
In this article, we explore this scenario and clarify the condition required for the successful 
Universe. 
Note that the essence of the scenario is the generation of chiral asymmetry 
carried mainly by the right-handed electrons without $B$-$L$ asymmetry 
and hence can be applied for other models beyond the SM than GUTs. 
%\\
%\underline{\it Generation and evolution of hypermagnetic fields}
\section{Generation and evolution of hypermagnetic fields}
First, we give an analytic explanation of how hyperMFs are generated at high temperature above the EW scale, 
which is consistent with recent numerical studies~\cite{Rogachevskii:2017uyc}. 
The basic equations to be solved are as follows. 
The evolution of the hypergauge fields (${\bm E}_Y, {\bm B}_Y$) 
in the comoving frame (with conformal time~$\tau$) is described by 
the Maxwell's equations,
\begin{equation}
\frac{d{\bm B}_Y}{d \tau} = -{\bm \nabla} \times {\bm E}_Y, \quad {\bm \nabla} \times {\bm B}_Y = {\bm J}_Y.   \label{max}
\end{equation}
Here, we  omit the displacement current $d{\bm E}_Y/d\tau$ since 
it is suppressed by the order of the amplitude of the fluid velocity ${\bm v}$ and large hyperelectric conductivity 
in the magnetohydrodynamic (MHD) approximations, 
which is appropriate to describe the dynamics of large-scale gauge fields we are interested in. 
The electric current consists of the Ohm's current and chiral magnetic current~\cite{Vilenkin:1980fu}, 
\begin{equation}
{\bm J}_Y = \sigma_Y ({\bm E}_Y+{\bm v}\times {\bm B}_Y) + \frac{2 \alpha_Y}{\pi} \mu_{5,Y} {\bm B}_Y. \label{cur}
\end{equation}
Here, $\sigma_Y\simeq 10^2 T_i$ is the hyperelectric conductivity~\cite{Baym:1997gq}, 
$\alpha_Y = g'^2/4\pi$ is the hyper-fine structure constant, 
and $\mu_{5,Y}=\sum_i (-1)^{\xi_i} y_i^2 \mu_i $ is the comoving chiral hyper-chemical potential. 
$T_i$ is the temperature where we define the scale factor $a(T)$ to be 1,  
$\xi_i$ is assigned for 0 for right-handed fermions and 1 for the left-handed fermions, 
and $y_i$ and $\mu_i$ are the hypercharge and comoving chemical potential of the fermion~$i$, respectively. 
The evolution of the chemical potential is determined by the anomaly equations. 
In the SM, since the sphalerons and Yukawa interactions except for the first-generation electron's one
are in equilibrium at high temperature $T\gtrsim 80 {\rm TeV}$~\cite{Campbell:1992jd}, 
the evolution for $\mu_{5,Y}$ is determined by the most weakly-coupled fermion, 
that is, the right-handed electrons~\cite{Joyce:1997uy,Giovannini:1997gp,Dvornikov:2012rk,Kamada:2016eeb},  
\begin{equation}
\frac{d \mu_{5,Y}}{d \tau} = c_1 \frac{6 y_{e_R}^2 \alpha_Y}{\pi T_i^2} \overline{ {\bm E}_Y\cdot {\bm B}_Y} - \Gamma_{\rm h \leftrightarrow ee} \mu_{5,Y}, \label{anom}
\end{equation}
where $c_1= \mu_{5,Y}/\mu_{e_R^1}$ is the ratio between the chiral and right-handed 
electron chemical potential and 
$\Gamma_{h\leftrightarrow ee}$ is the chirality flip rate
of the first-generation electron Yukawa interaction. 
The overline represents the volume average. 
The evolution of the fluid velocity ${\bm v}$ is, in principle, described by the Navier-Stokes
equation, which is hard to solve. 
However, numerical studies showed that the velocity fields are 
emerged from vanishing initial conditions 
due to the Lorentz force and reach at an equilibrium to the hyperMFs immediately~\cite{Rogachevskii:2017uyc,Banerjee:2004df}. 
Thus, we assume here that the velocity fields obtain the comparable 
strength to the hyperMFs with a similar coherent length, 
${\cal E}_M =  a(T)^{-4} \overline{|{\bm B}_Y|^2}/2 = \gamma^{-2} {\cal E}_K = \gamma^{-2} \rho \overline{|{\bm v}|^2}/2$  with $\rho \equiv (30/\pi^2) g_* T^4$ 
being the energy density of the Universe. 
The number of relativistic degrees of freedom $g_*$ is taken to be 106.75 for the SM.   
The ratio between the velocity and MF strength $\gamma$ is found to be ${\cal O}(0.01) - {\cal O}(1)$~\cite{Rogachevskii:2017uyc,Banerjee:2004df}. 
From Eqs.~\eqref{max}, \eqref{cur}, and \eqref{anom}, we can remove the electric fields as
\begin{align}
\frac{d {\bm B}_Y}{d\tau} =&\frac{1}{\sigma_Y}\left( {\bm \nabla}^2 {\bm B}_Y + \frac{2 \alpha_Y}{\pi} \mu_{5,Y} {\bm \nabla} \times {\bm B}_Y \right)+{\bm \nabla} \times ( {\bm v} \times {\bm B}_Y),  \label{bevo}\\
\frac{d \mu_{5,Y}}{d\tau} =& \frac{6 c_1 y_{e_R}^2\alpha_Y}{\pi T_i^2 \sigma_Y} \overline{{\bm B}_Y\cdot ({\bm \nabla} \times {\bm B}_Y)} \notag \\  &-\left( \frac{12 c_1 y_{e_R}^2\alpha_Y^2}{\pi^2 T_i^2\sigma_Y } \overline{{\bm B}_Y^2} +\Gamma_{h\rightarrow ee}  \right) \mu_{5,Y} . \label{mu5}
\end{align}
From these equations, we can see that there is a conserved quantity, 
$\mu_{5,Y}+3 c_1 y_{e_R}^2\alpha_Y h /(\pi T_i^2)$ 
in the limit where the Yukawa interaction $\Gamma_{h\rightarrow ee}$ is negligible.  Here, 
$h \equiv (1/V)\int_V d^3 x {\bm Y} \cdot {\bm B}_Y$ is the hypermagnetic helicity density 
with ${\bm Y}$ being the hypercharge
vector potential. 

Let us investigate the evolution of the system with initial conditions with a large $\mu_{5,Y}=\mu_{5,Y}^i$ 
and tiny  ${\bm B}_Y$ and ${\bm v}$. 
When the hyperMFs are small enough, $\mu_{5,Y}$ is effectively constant 
and the last term in Eq.~\eqref{bevo} is negligible. 
Then the equation for the circular polarization modes of the hyperMFs is
\begin{equation}
\frac{d B_k^{Y\pm}}{d \tau} = -\frac{k}{\sigma_Y}\left(k \mp \frac{2 \alpha_Y}{\pi} \mu_{5,Y}^i\right) B_k^{Y\pm}. 
\end{equation}
If $\mu_{5,Y}^i>0 (<0)$, the +(-) mode feels instability for smaller $k$ while the -(+) 
mode does not. 
The difference between two circular polarization mode is the hypermagnetic helicity
and hence the amplified hyperMFs are maximally helical. 
Since the resultant baryon asymmetry is positive for positive $\mu_{5,Y}^i$, 
we hereafter focus on the case with $\mu_{5,Y}^i>0$. 
The most unstable mode is at $k_c=\alpha_Y \mu_{5,Y}^i/\pi$ and evolves as
$B_k^{Y+} \propto \exp[k_c^2 \tau /\sigma_Y]$. 
Thus at the time $\tau \simeq \sigma_Y/k_c^2 \equiv \tau_c, $
the instabilities start to grow. 
This corresponds to the temperature 
\begin{align}
T_c \simeq & 6.8  \times 10^6 {\rm GeV} \left(\frac{\alpha_Y}{10^{-2}}\right)^2 \left(\frac{\sigma_Y}{10^2 T_i}\right)^{-1}  \notag \\
&\times \left(\frac{g_*}{106.75}\right)^{-1/2} \left(\frac{\mu^i_{5,Y}/T_i}{10^{-2}}\right)^2. 
\end{align}
If $|\mu_{5,Y}^i|/T_i \gtrsim 10^{-3}$, the hyperMFs are amplified at $T\gtrsim 80 {\rm TeV}$, 
before the electron Yukawa interaction gets effective and the $\mu_{5,Y}$ starts to decay. 

As the hyperMFs grow, the Lorentz force drives the velocity fields up to $v \sim \gamma B_Y$ in a short period 
$\left(v \equiv \overline{|{\bm v}|^2}^{1/2}, B_Y \equiv \overline{|{\bm B}_Y|^2}^{1/2}\right)$. 
Since the velocity fields erase the small scale structure, the effects of the fluid velocity on the evolution of hyperMFs 
are no longer negligible when the eddy turnover scale 
$\lambda_{\rm et} = v \tau$ reaches at the instability scale $\lambda_c \equiv 2 \pi/k_c$. 
This happens when the hyperMFs evolve up to $B_Y \simeq (\pi^2 g_*/30)^{1/2} 2 \alpha_Y \mu_{5,Y}^i T_i^2/(\gamma \sigma_Y)$. 
For $c_1 \simeq {\cal O}(1)$, the hypermagnetic helicity is still smaller than the chiral chemical potential, $\mu_{5,Y} \gg 3 c_1 y_{e_R}^2 \alpha_Y h /(\pi T_i^2)$, and the hyperMFs still continue to be amplified, with the comoving coherent length satisfying
$\lambda_Y \simeq v \tau \simeq \gamma B_Y \tau/\sqrt{\rho}$~\cite{Rogachevskii:2017uyc}. 
By estimating the amplification time scale as 
$\tau \sim \sigma_Y \lambda_Y /4 \alpha_Y  \mu_{5,Y}$, 
we can have the mean field strength and coherent length at given time $\tau$. 

The amplification of the hyperMFs terminates when the hypermagnetic helicity is saturated, 
$3 c_1 y_{e_R}^2 \alpha_Y h /(\pi T_i^2)\simeq \mu_{5,Y}^i$. 
The resultant physical hyperMF strength and coherent length at the saturation are evaluated as 
\begin{align}
B_Y^{\rm phys}(T_{\rm s})\simeq &  1.4 \times 10^{10} {\rm GeV}^2 c_1^2 \left(\frac{\gamma}{10^{-2}}\right)^{-5} \left(\frac{\alpha_Y}{10^{-2}}\right)^{9} \notag \\
& \times \left(\frac{\sigma_Y/T_i}{10^2}\right)^{-7} \left(\frac{\mu_{5,Y}^i/T_i}{10^{-2}}\right)^{5} \left(\frac{g_*}{106.75}\right)^{3/2}, \\
\lambda_Y^{\rm phys}(T_{\rm s}) \simeq & 0.48 \ {\rm GeV}^{-1} c_1^{-2}  \left(\frac{\gamma}{10^{-2}}\right)^{4} \left(\frac{\alpha_Y}{10^{-2}}\right)^{-7} \notag \\ 
& \times  \left(\frac{\sigma_Y/T_i}{10^2}\right)^{5} \left(\frac{\mu_{5,Y}^i/T_i}{10^{-2}}\right)^{-3} \left(\frac{g_*}{106.75}\right)^{-3/2}, 
\end{align}
at the temperature 
\begin{align}
T_{\rm s} \simeq &2.4 \times 10^6 {\rm GeV} c_1  \left(\frac{\gamma}{10^{-2}}\right)^{-2} \left(\frac{\alpha_Y}{10^{-2}}\right)^{4} \notag \\
&\times \left(\frac{\sigma_Y/T_i}{10^2}\right)^{-3} \left(\frac{\mu_{5,Y}^i/T_i}{10^{-2}}\right)^{2} \left(\frac{g_*}{106.75}\right)^{1/2}. 
\end{align}
After the saturation, the hyperMFs evolve according to the inverse cascade law, 
$B_Y \propto \tau^{-1/3}, \lambda_Y \propto \tau^{2/3}$ \cite{Rogachevskii:2017uyc,Banerjee:2004df}
supported by the velocity fields, while the chiral asymmetry is erased so that Eq.~\eqref{mu5} reaches to the 
equilibration, $d\mu_{5,Y}/d\tau \simeq 0$~\cite{Giovannini:1997gp,Dvornikov:2012rk,Kamada:2016eeb}. 
At the temperature $T$ during radiation domination before the electroweak symmetry breaking, 
the physical properties of the hyperMFs are given by 
\begin{gather}
B_Y^{\rm phys}(T) \simeq 0.82 {\rm GeV}^2 \ \left(\frac{g_{*s}(T)}{g_{*s}(T_{\rm s})}\right)^{7/9} c_1^{-1/3}  \left(\frac{\gamma}{10^{-2}}\right)^{-1/3} \notag \\
\times \left(\frac{\alpha_Y}{10^{-2}}\right)^{-1/3}\left(\frac{\mu_{5,Y}^i/T_i}{10^{-2}}\right)^{1/3} \left(\frac{g_*}{106.75}\right)^{1/3} \left(\frac{T}{10^2 {\rm GeV}}\right)^{7/3}, \label{BT}\\
\lambda_Y^{\rm phys}(T) \simeq 9.8 \times 10^6 {\rm GeV}^{-1} \ \left(\frac{g_{*s}(T_{\rm s})}{g_{*s}(T)}\right)^{5/9} c_1^{-1/3}  \left(\frac{\gamma}{10^{-2}}\right)^{2/3} \notag \\
\times \left(\frac{\alpha_Y}{10^{-2}}\right)^{-1/3}\left(\frac{\mu_{5,Y}^i/T_i}{10^{-2}}\right)^{1/3} \left(\frac{g_*}{106.75}\right)^{-2/3} \left(\frac{T}{10^2 {\rm GeV}}\right)^{-5/3}. \label{lamT}
\end{gather}
The MFs continue to evolve until today and exist in the intergalactic void with the properties
\begin{align}
B_{\rm phys}^0 \simeq & 9.9 \times 10^{-16} {\rm G} \ c_1^{-1/3}  \left(\frac{\gamma}{10^{-2}}\right)^{-1/3} \left(\frac{\alpha_Y}{10^{-2}}\right)^{-1/3} \notag \\
& \times \left(\frac{\mu_{5,Y}^i/T_i}{10^{-2}}\right)^{1/3} \left(\frac{g_*}{106.75}\right)^{1/3}, \\
\lambda_{\rm phys}^0  \simeq & 6.9  \times 10^{-3} {\rm pc} \ c_1^{-1/3}  \left(\frac{\gamma}{10^{-2}}\right)^{2/3} \left(\frac{\alpha_Y}{10^{-2}}\right)^{-1/3} \notag \\
&\times \left(\frac{\mu_{5,Y}^i/T_i}{10^{-2}}\right)^{1/3} \left(\frac{g_*}{106.75}\right)^{-2/3}. 
\end{align}
These intergalactic MFs with positive helicities are the smoking-gun of that scenario, 
as argued in Ref.~\cite{Kamada:2016eeb}. 
%\\
%\underline{\it (Re)generation of baryon asymmetry}
\section{(Re)generation of baryon asymmetry}
Equation~\eqref{mu5} suggests that the chiral asymmetry (as well as the baryon asymmetry) will not be completely washed out
in the presence of hypermagnetic helicity but reach at the equilibration, $d\mu_{5,Y}/d\tau \simeq 0$~\cite{Giovannini:1997gp,Dvornikov:2012rk,Kamada:2016eeb},  
\begin{equation}
\mu_{5,Y} \simeq \frac{12 \pi  c_1 y_{e_R}^2 \alpha_Y B_Y^2/\lambda_Y}{24 c_1 y_{e_R}^2 \alpha_Y^2 B_Y^2/\pi +  T_i^2 \sigma_Y \Gamma_{h \rightarrow ee}}.  
\end{equation}
Note that for the maximally helical fields with a positive helicity, it is approximated as $\overline{{\bm B}_Y \cdot ({\bm \nabla}\times {\bm B})} \simeq 2 \pi B_Y^2/\lambda_Y$. 
Moreover, when the electroweak symmetry breaking takes place, the hypermagnetic helicity is transferred to the 
(electro)magnetic helicity, which gives a nonzero contribution to the anomaly equation for the $B$+$L$ asymmetry. 
This effect has been studied in detail in Ref.~\cite{Kamada:2016eeb}, which shows that 
in the SM crossover with the 125 GeV Higgs boson 
the effect lasts for a while 
after the freezeout of the EW sphalerons and hence the $B$+$L$ asymmetry is not washed out completely. 
The resultant baryon and lepton asymmetry of the Universe today is evaluated as
\begin{equation}
\eta_B^0 \equiv \left.\frac{n_B}{s}\right|_{\rm today}  \simeq \frac{17}{37} \left[(g^2+g'^2)\frac{f(\theta_W,T){\cal S}}{\gamma_{\rm w, sph}}\right]_{T=135{\rm GeV}}
\end{equation}
with~\cite{Jimenez:2017cdr}
\begin{align}
\left.f(\theta_W,T)\right|_{T=135 {\rm GeV}}& \equiv -T \frac{d \theta_W}{dT} \sin(2\theta_W) \notag \\
& \simeq (5 \times 10^{-4} \cdots  0.3), \\
{\cal S} &\equiv \frac{H}{sT} \frac{\lambda_Y^{\rm phys} (B_Y^{\rm phys})^2}{16 \pi^3}, \\
\gamma_{\rm w, sph} &= \exp \left[-147.7+107.9 \left(\frac{T}{130 {\rm GeV}}\right)\right], 
\end{align}
where $g$ and $g'$ are the SM SU(2)$_L$ and U(1)$_Y$ gauge couplings, 
$s$ is the entropy density, $n_B$ is the baryon number density, $\theta_W$ is the temperature-dependent effective 
weak mixing angle, 
and $H$ is the Hubble parameter. 
The uncertainty in $f(\theta_W,T)$ comes from the errors in the temperature dependence of 
the weak mixing angle in the EW crossover found in the one-loop calculation and lattice calculations~\cite{Kamada:2016eeb,Kajantie:1996qd}.\footnote{
Note that the  uncertainty in $\left.f(\theta_W,T)\right|_{T=135 {\rm GeV}}$ is a conservative one, 
which includes the fitting function whose fit is not so good at $T\lesssim 150$ GeV (Fitting function B of Fig. 2 in the second article of Ref.~\cite{Kamada:2016eeb}). If we omit 
the fitting function, we have $\left.f(\theta_W,T)\right|_{T=135 {\rm GeV}} \simeq (0.04 ... 0.3)$. } 
Since the hyperMFs generated from the chiral instability discussed in the above 
are maximally helical and the mechanism in Refs.~\cite{Giovannini:1997gp, Dvornikov:2012rk,Kamada:2016eeb} work. 
From Eqs.~\eqref{BT} and \eqref{lamT}, the resultant baryon asymmetry today in terms of the initial chiral asymmetry is
calculated as
\begin{align}
\eta_B^0 \simeq &4.0 \times 10^{-5} c_1^{-1}  \left(\frac{\alpha_Y}{10^{-2}}\right)^{-1}\left(\frac{\mu_{5,Y}^i/T_i}{10^{-2}}\right)  f(\theta_W, T), 
\end{align}
which is the main result of this article. 
Therefore, although precise evaluations of the 
temperature dependence of the weak mixing angle are needed for the quantitatively precise estimate, 
the helical hyperMFs generated by the chiral instability can be responsible for the present BAU 
$\eta_B^0 \sim 10^{-10}$ if the initial chiral asymmetry is $\mu_{5,Y}^i/T_i \sim 10^{-3}$ 
and $f(\theta_W, T=135 {\rm GeV})\sim 10^{-4}$ with $c_1 = {\cal O}(1)$. 
Note that this predicts slightly large baryon asymmetry, but it is not problematic since 
for $\mu_{5,Y}^i/T_i \sim 10^{-3}$ the generation of hypermagnetic fields might not be saturated 
before $T\sim 80$ TeV when the chirality flip interaction becomes strong, and hence the resultant magnetic fields properties
can be slightly smaller so that they are appropriate for the present BAU. 
On the other hand, for  $\mu_{5,Y}^i/T_i \ll 10^{-3}$, the chiral instability of hyperMFs do not occur at all. 
Consequently, even if  $f(\theta_W, T=135 {\rm GeV})\gg 10^{-4}$ the present BAU cannot be explained. 
It should be also noted that we suffer from baryon overproduction  if $f(\theta_W, T=135 {\rm GeV})\gg 10^{-4}$ and $\mu_{5,Y}^i/T_i \gtrsim 10^{-3}$ for $c_1 = {\cal O}(1)$. 
But if $c_1$ is appropriately large due to the nature of the chiral asymmetry generation mechanism, 
$f(\theta_W, T=135 {\rm GeV})\gg 10^{-4}$ and $\mu_{5,Y}^i/T_i \gg10^{-3}$ can be accommodate to the present BAU. 

%\\
%\underline{\it Possible source of the chiral asymmetry}
\section{Possible source of the chiral asymmetry}
Finally, let us give a discussion on the possible origin of such large chiral asymmetry 
mainly carried by the right-handed electrons. 
In the standard SU(5) GUT baryogenesis~\cite{Yoshimura:1978ex}, 
thermally produced GUT bosons decay into quarks and leptons with nonvanishing $B$+$L$ asymmetry.  
However, 
the thermal GUT symmetry breaking that would occur in that case suffers from the monopole problem. 
This problem can be evaded if the GUT Higgs bosons in the ${\bm 5}$ representation of SU(5) 
are produced by the preheating~\cite{Kolb:1996jt}. 
In that case, large chiral asymmetries can be generated. 
However, if we identify the Higgs boson (${\bm 5}$) is responsible for the EW symmetry breaking, 
it mainly decays into the third or second generation fermions through the Yukawa interaction 
and little right-handed electron asymmetry 
is generated. As a result $c_1 = \mu_{5,Y}/\mu_{e_R^1}$ is extremely large and hence the hyperMFs
as well as the resultant baryon asymmetry is highly suppressed.  

One way to overcome these difficulties is to consider another Higgs field in the ${\bm 5}$ representation of SU(5) GUT 
that is not related to the EW symmetry breaking but mainly couples to the first generation fermions. 
Imagine they once dominate the energy density of the Universe through, e.g., instant preheating~\cite{Felder:1998vq} 
and eventually decay into $e_R^1 u_R^1$ and ${\bar Q}_L^1 {\bar Q}_L^1$ pairs
after they become nonrelativistic.  Then a large chiral chemical potential mainly carried by right-handed electrons  can be 
generated. In that case, while the asymmetry of the right-handed electrons are unchanged for a while, 
the asymmetries are rearranged to other fermions through the Yukawa interaction and sphaleron processes immediately 
so that we obtain $c_1=\mu_{5,Y}/\mu_{e_R^1} = 553/481$. 
The (physical) chemical potential at the time of decay is evaluated as 
\begin{equation}
\frac{\mu_{5,Y}^{\rm phys}}{T_{\rm dec}} = \frac{\mu_{5,Y}^i}{T_i} = \frac{\pi^2 g_*}{5}c_1 \epsilon \frac{T_{\rm dec}}{m_X},
\end{equation}
where $\epsilon$ is the net right-handed electron asymmetry produced by a single Higgs-anti Higgs pair, 
$T_{\rm dec}$ is the Higgs decay temperature, and $m_X$ is the mass of the Higgs field. 
Therefore, if e.g., the CP violation in GUT allows $\epsilon \simeq 10^{-3}$ and the decay temperature of the Higgs boson is tuned 
to be  $T_R/m_X \sim 10^{-2}$, the chiral asymmetry ideal for the present BAU can be generated.  
%\\
%\underline{\it Conclusion}
\section{Conclusion}
In this article, we discussed the possibility for baryogenesis models without $B$-$L$ generation 
such as GUT baryogenesis 
to be indirectly responsible for the present BAU. 
It is usually considered that $B$+$L$ asymmetry is washed out completely by the EW sphalerons 
and no asymmetry would remain. 
In the scenario discussed here, the washout by the EW sphalerons are evaded 
by the mechanism that the asymmetry is first transferred to the hypermagnetic helicity 
which in insensitive to the EW sphaleron. 
Baryon asymmetry is provided by the hypermagnetic helicity decay that is effective until shortly after the EW sphaleron 
freezeout~\cite{Kamada:2016eeb}. 

There are several difficulties for the realistic model building. 
The efficiency is not as much as 100\% and relatively large initial asymmetry is required, $\mu_{5,Y}/T_i \simeq 10^{-3}$.  
In addition, the asymmetry should be carried mainly by the right-handed electrons. 
Nevertheless, our findings here opened a new direction in the study of the BAU. 
Further studies on the realistic model building for the generation of the $B$+$L$ asymmetry 
as well as the determination of the temperature dependence of the weak mixing angle are required. 

\section*{Acknowledgements}
K.K. is grateful to R.~Jinno, T.~Kuwahara, A.~J.~Long, 
T.~Vachaspati, M.~Yamaguchi and J.~Yokoyama
for useful comments and discussions.  
The work of K.K. was supported by Institute for Basic Science (IBS) 
under the Project Code IBS-R018-D1, and by the Department of Energy (DOE) 
under Grant No. DE-SC0013605.

\end{document}